\documentclass[sigconf,natbib=false]{acmart}

\setcopyright{none}
\acmYear{2025}
\acmConference[SC Workshops '25]{Workshops of the International Conference for High Performance Computing, Networking, Storage and Analysis}{November 16--21, 2025}{St Louis, MO, USA}
\acmBooktitle{Workshops of the International Conference for High Performance Computing, Networking, Storage and Analysis (SC Workshops '25), November 16--21, 2025, St Louis, MO, USA}
\acmDOI{}
\acmISBN{}

\usepackage[T1]{fontenc} %
\usepackage[utf8]{inputenc} %
\usepackage[english]{babel} %
\usepackage{amsmath,amsthm,amsfonts}
\usepackage{bm} %
\usepackage{comment} %
\usepackage{microtype}

\usepackage[backend=biber, style=acmnumeric,
    sortlocale=en_US,
    sortcites=true,
    url=true,
    eprint=true,
    minnames=1,
    maxnames=6
]{biblatex}
\AtEveryBibitem{%
    \clearname{editor}%
    \clearfield{series}%
    \clearfield{isbn}%
    \clearfield{issn}%
    \clearfield{day}%
    \clearfield{month}%
    \clearfield{place}%
    \clearlist{location}%
    \clearfield{pages}%
    \clearfield{volume}%
    \clearfield{number}%
    \ifentrytype{software}{%
    }{%
        \clearfield{url}%
        \clearfield{urlyear}%
        \clearfield{urldate}%
    }%
}
\DeclareSourcemap{
    \maps[datatype=bibtex]{
        \map{
            \step[fieldsource=doi, match=\regexp{.+/arXiv\..+}, final]
            \step[fieldsource=eprint, match=\regexp{.+}, final]
            \step[fieldsource=archiveprefix, match=\regexp{.+}, final]
            \step[fieldset=doi, null]
        }
        \map{
            \step[fieldsource=doi, match=\regexp{.+}, final]
            \step[fieldset=eprint, null]
            \step[fieldset=archiveprefix, null]
        }
    }
}
\usepackage{csquotes} %
\usepackage[inline]{enumitem} %
\usepackage{tabularx} %
\usepackage{booktabs} %
\usepackage[capitalize,nameinlink]{cleveref}
\usepackage{xcolor}
\usepackage{graphicx}
\usepackage{stmaryrd}
\usepackage{mathtools}

\usepackage{wrapfig} %
\usepackage{float}  %
\usepackage{placeins} %
\graphicspath{ {figures/} }
\usepackage{fontawesome5}
\faStyle{regular}
\usepackage{xspace}
\usepackage[noEnd]{algpseudocodex}

\usepackage{fnpct} %
\usepackage{thmtools}
\declaretheoremstyle[
    bodyfont=\itshape,%
]{example-style}
\declaretheorem[
    name=Example,%
    style=example-style,%
    numbered=unless unique,%
]{example}

\usepackage{siunitx}
\usepackage{listings}
\usepackage{tikz}
\usetikzlibrary{fit}
\usepackage{tikzpagenodes}
\usepackage{atbegshi}
\usepackage{refcount}
\usepackage{braket}
\usepackage{quantikz}

\crefname{listing}{Code}{Code}
\Crefname{listing}{Code}{Code}
\crefname{section}{Sec.}{Sec.}
\Crefname{section}{Sec.}{Sec.}
\crefname{example}{Ex.}{Ex.}
\Crefname{example}{Ex.}{Ex.}

\newcolumntype{R}{>{\raggedleft\arraybackslash}X}
\newcolumntype{C}{>{\centering\arraybackslash}X}

\newcommand{\emptylines}[1]{%
    \newcount\foo%
    \foo=0%
    \loop%
    \phantom{}\\%
    \advance\foo 1%
    \ifnum \foo<#1%
    \repeat%
}

\definecolor{TUM_blue}{RGB}{0,101,189}
\colorlet{TUM_black}{black}
\colorlet{TUM_white}{white}
\definecolor{TUM_darkblue}{RGB}{0,82,147}
\colorlet{TUM_darkblue100}{TUM_darkblue}
\colorlet{TUM_darkblue80}{TUM_darkblue100!80}
\colorlet{TUM_darkblue50}{TUM_darkblue100!50}
\colorlet{TUM_darkblue20}{TUM_darkblue100!20}
\definecolor{TUM_verydarkblue}{RGB}{0,51,89}
\colorlet{TUM_verydarkblue100}{TUM_verydarkblue}
\colorlet{TUM_verydarkblue80}{TUM_verydarkblue100!80}
\colorlet{TUM_verydarkblue50}{TUM_verydarkblue100!50}
\colorlet{TUM_verydarkblue20}{TUM_verydarkblue100!20}
\colorlet{TUM_darkgrey}{TUM_black!80}
\colorlet{TUM_grey}{TUM_black!50}
\colorlet{TUM_lightgrey}{TUM_black!20}
\definecolor{TUM_beige}{RGB}{218,215,203}
\definecolor{TUM_orange}{RGB}{227,114,34}
\definecolor{TUM_green}{RGB}{162,173,0}
\definecolor{TUM_verylightblue}{RGB}{152,198,234}
\definecolor{TUM_lightblue}{RGB}{100,160,200}

\newcommand{\ie}{i.\,e.\nolinebreak\@\xspace}
\newcommand{\eg}{e.\,g.\nolinebreak\@\xspace}
\newcommand{\Eg}{E.\,g.\nolinebreak\@\xspace}
\renewcommand{\phi}{\varphi}

\lstdefinelanguage{qasm}{
    keywords=[1]{include,creg,qreg},
    keywords=[2]{h,cx,x,y,z,s,sdg,t,tdg,rx,ry,rz,measure,barrier,reset},
    otherkeywords={[,],->},
    string=[b]",
    comment=[l]//,
    morecomment=[l]OPENQASM
}
\lstdefinelanguage{qir}{
    language=llvm,
    morekeywords=[2]{%
        ptr
    },
    morekeywords=[3]{%
        @__quantum__qis__h__body,%
        @__quantum__qis__cnot__body,%
        @__quantum__qis__mz__body,%
        @__quantum__qis__m__body,%
        @__quantum__rt__array_get_element_ptr_1d,%
        @__quantum__rt__qubit_allocate_array,%
        @__quantum__rt__array_create_1d,%
        @__quantum__rt__qubit_release_array,%
        @__quantum__rt__result_update_reference_count,%
        @__quantum__rt__result_update_reference_count,%
        @__quantum__rt__array_update_reference_count%
    },
}
\lstdefinelanguage{qir++}{
    language=C++,
    morekeywords=[2]{%
        size_t,%
        uintptr_t,%
        int32_t,%
        int8_t,%
        int64_t,%
        uint64_t,%
        QIR_DD_Backend%
    },
    morekeywords=[3]{%
        __quantum__qis__h__body,%
        __quantum__qis__rx__body,%
        __quantum__qis__cnot__body,%
        __quantum__qis__mz__body,%
        __quantum__qis__m__body,%
        __quantum__rt__array_get_element_ptr_1d,%
        __quantum__rt__qubit_allocate,%
        __quantum__rt__qubit_allocate_array,%
        __quantum__rt__array_create_1d,%
        __quantum__rt__array_update_reference_count,%
        translateAddresses,%
        enlargeState,%
        qAlloc,%
        getInstance%
    },
}

\usepackage{flushend} %

\begin{document}

    \title{Towards Supporting QIR}
    \subtitle{Steps for Adopting the Quantum Intermediate Representation}

    \author{Yannick Stade}
    \email{yannick.stade@tum.de}
    \affiliation{%
        \institution{Technical University of Munich}
        \country{Germany}
    }
    \author{Lukas Burgholzer}
    \email{lukas.burgholzer@tum.de}
    \affiliation{%
        \institution{Technical University of Munich}
        \country{}
    }
    \affiliation{%
        \institution{Munich Quantum Software Company}
        \country{Germany}
    }
    \author{Robert Wille}
    \email{robert.wille@tum.de}
    \affiliation{%
        \institution{Technical University of Munich}
        \country{}
    }
    \affiliation{%
        \institution{Munich Quantum Software Company}
        \country{Germany}
    }

    \hypersetup{ 
        pdftitle={Towards Supporting QIR: Steps for Adopting the Quantum Intermediate Representation},
        pdfauthor={
            Yannick Stade,
            Lukas Burgholzer,
            Robert Wille
        }
    }

    \begin{abstract}
        Intermediate representations (IRs) play a crucial role in the software stack of a quantum computer to facilitate efficient optimizations for executing an application on hardware.
        One of those IRs is the \emph{Quantum Intermediate Representation}~(QIR), which builds on the classical LLVM compiler infrastructure.
        In this article, we outline different approaches to how QIR can be adopted.
        This exploration culminates in a demonstration of what it takes to turn an existing quantum circuit simulator into a QIR runtime and that such a transition is less daunting than it might seem at first.
        We further show that switching to QIR does not entail any performance deficits compared to the original simulator.
        On the contrary, the presented steps effortlessly allow adding support for arbitrary classical control flow to any classical simulator.
        We conclude with an outlook on future directions using QIR.
        The implemented QIR runtime is available under \mbox{\url{https://github.com/munich-quantum-toolkit/core}}.
    \end{abstract}

    \ccsdesc[500]{Hardware~Quantum computation}

    \keywords{quantum computing, intermediate representation, compiler, runtime, software stack}

    \maketitle

    \section{Introduction}\label{sec:introduction}

    Since the famous results of Shor~\cite{shorAlgorithmsQuantumComputation1994}, the promise holds that quantum computers can offer an exponential speedup over classical computers for specific problems.
    Many applications~\mbox{\cite{stamatopoulosOptionPricingUsing2020,peruzzoVariationalEigenvalueSolver2014,harwoodFormulatingSolvingRouting2021,zoufalQuantumGenerativeAdversarial2019,groverQuantumMechanicsHelps1997}} have since been developed to exploit this potential computational power~\cite{bernsteinQuantumComplexityTheory1997,boyerTightBoundsQuantum1999}.
    However, physical implementations of quantum computers still need to catch up and are far from being useful for the aforementioned applications, with only a few realizations of toy examples on real hardware devices reported, \eg,~\cite{vandersypenExperimentalRealizationShors2001,bluvsteinLogicalQuantumProcessor2023}.
    Nevertheless, the development of real quantum computers has seen rapid progress over the past years.
    Breakthroughs in different technologies, such as superconducting qubits~\cite{googlequantumaiSuppressingQuantumErrors2023,kimEvidenceUtilityQuantum2023}, trapped ions~\cite{mosesRaceTrackTrappedIon2023}, and neutral atoms~\cite{bluvsteinLogicalQuantumProcessor2023,manetschTweezerArray61002024}, have demonstrated significant advancements in the number of qubits and their reliability.

    However, hardware is only one part of the story: Successfully realizing applications on quantum computers requires a sophisticated stack of software tools to transform a classical problem description into a quantum solution, to optimize the resulting (hybrid) quantum-classical program, and, finally, to execute it~\cite{burgholzer2025mqss,burgholzerBuildingEfficientSoftware2024}.
    Since real quantum computers are not capable of executing arbitrary instructions directly, a compiler with various transformation and optimization passes is needed~\cite{liTacklingQubitMapping2019,willeMQTQMAPEfficient2023,schoenbergerUsingBooleanSatisfiability2024,stadeAbstractModelEfficient2024,quetschlichMQTPredictorAutomatic2025,willeMQTHandbookSummary2024}.
    These passes are essential to minimize the overhead introduced by the compilation process and to maintain a high fidelity of the resulting quantum program.
    Many of the problems involved in this process are computationally hard to solve, making the development of such a compiler or a general compiler infrastructure challenging.

    Intermediate representations are a key component of any compiler, enabling the seamless combination of various passes and transformations.
    Huge parts of the quantum computing ecosystem have come to rely on the OpenQASM format, proposed initially as version~2~\cite{crossOpenQuantumAssembly2017} and refined in version~3~\cite{crossOpenQASMBroaderDeeper2022}, by IBM.
    The format started as a low-level quantum assembly language and has since evolved into a more general-purpose quantum programming language, adding classical elements.

    In contrast, the \emph{Quantum Intermediate Representation}~(QIR), initially proposed by \mbox{Microsoft}~\cite{QIRSpec2021}, takes a different approach.
    It aims to reuse as much as possible of the existing classical compiler infrastructure, namely the LLVM compiler framework~\cite{lattnerLLVMCompilationFramework2004}, and augment it with quantum instructions.
    Despite its promises, the adoption of QIR in the quantum computing ecosystem remains low, especially outside the realm of industrial players like \mbox{Microsoft}, \mbox{Nvidia}, or \mbox{Xanadu} that have the resources and the expertise to develop and maintain such a compiler infrastructure.

    In this article, we aim to shine a light on the challenges and opportunities that come with adopting QIR in the quantum computing software ecosystem.
    This should help readers to develop a better understanding of the various options to support QIR in their software tools.
    In particular, we detail the steps that are necessary to turn an existing quantum circuit simulator into a QIR runtime.
    In this regard, we show how to implement the LLVM IR interface defined by QIR for a simulator that is written in C++.
    Meanwhile, we elaborate on the different resource management strategies supported by QIR and how to deal with them in a classical runtime for quantum computing.
    The complete code is publicly available in open-source as part of the \emph{Munich Quantum Toolkit}~(MQT,~\cite{willeMQTHandbookSummary2024}) under \url{https://github.com/munich-quantum-toolkit/core}.

    \AtBeginShipout{\ifnum
        \value{page}=\getpagerefnumber{fig:code}
        \AtBeginShipoutUpperLeft{%
            \begin{tikzpicture}[remember picture,overlay,fill=TUM_darkblue20,inner sep=0,every node/.style={fill=TUM_darkblue20,inner sep=0}]
                \node[fit=(qreg)] {};
                \node[fit=(creg)] {};
                \node[fit=(h)] {};
                \node[fit=(cx)] {};
                \node[fit=(measure)] {};
                \node[fit=(qreg begin)(qreg end)] {};
                \node[fit=(creg begin)(creg end)] {};
                \node[fit=(h begin)(h end)] {};
                \node[fit=(cx begin)(cx end)] {};
                \node[fit=(measure begin)(measure end)] {};
                \fill (qreg.north east) .. controls ++(1,0) and ++(-1,0) .. (qreg begin.north west) -- (qreg end.south west) .. controls ++(-1,0) and ++(1,0) .. (qreg.south east) -- cycle;
                \fill (creg.north east) .. controls ++(1,0) and ++(-1,0) .. (creg begin.north west) -- (creg end.south west) .. controls ++(-1,0) and ++(1,0) .. (creg.south east) -- cycle;
                \fill (h.north east) .. controls ++(1,0) and ++(-1,0) .. (h begin.north west) -- (h end.south west) .. controls ++(-1,0) and ++(1,0) .. (h.south east) -- cycle;
                \fill (cx.north east) .. controls ++(1,0) and ++(-1,0) .. (cx begin.north west) -- (cx end.south west) .. controls ++(-1,0) and ++(1,0) .. (cx.south east) -- cycle;
                \fill (measure.north east) .. controls ++(1,0) and ++(-1,0) .. (measure begin.north west) -- (measure end.south west) .. controls ++(-1,0) and ++(1,0) .. (measure.south east) -- cycle;
            \end{tikzpicture}
        }
    \fi}
    \begin{figure*}[htp]
        \tikzset{
            loc/.style={font=\scriptsize\ttfamily,anchor=base west,inner xsep=2pt,inner ysep=.5pt}
        }
        \def\qasmline#1{%
            \begin{tikzpicture}[remember picture,overlay]
                \node[loc,minimum width=\linewidth] (#1) at (current bounding box.base -| current page text area.west) {\vphantom{dj}};
            \end{tikzpicture}
        }
        \def\qirline#1{%
            \begin{tikzpicture}[remember picture,overlay]
                \node[loc,minimum width=\linewidth] (#1) at (current bounding box.base -| qir east) {\vphantom{dj}};
            \end{tikzpicture}
        }
        \hrule width \linewidth \relax\vspace{-2pt}
        \begin{minipage}[t]{.2\textwidth}
            \begin{lstlisting}[language=qasm,frame=none]
OPENQASM 2.0;
include "qelib1.inc";
qreg q[2]; (@\qasmline{qreg}\vspace{1pt}@)
creg c[2]; (@\qasmline{creg}\vspace{1pt}@)
h q[0]; (@\qasmline{h}\vspace{1pt}@)
cx q[0], q[1]; (@\qasmline{cx}\vspace{1pt}@)
measure q -> c; (@\qasmline{measure}\vspace{1pt}@)
            \end{lstlisting}
            \vspace{4ex}
            \begin{center}
                \begin{quantikz}
                    \lstick{$\ket{0}$} & \gate{H} & \ctrl{1} & \meter{} \\
                    \lstick{$\ket{0}$} & \qw       & \targ{}    & \meter{}
                \end{quantikz}
            \end{center}
        \end{minipage}
        \hfill
        \begin{minipage}[t]{.65\linewidth}
            \tikz[remember picture,overlay] \coordinate (qir east);
            \def\comment#1{}
            \begin{lstlisting}[language=qir,frame=none]
(@\hspace{1cm}\tikz[overlay] \node[anchor=base,yshift=-1pt] {{\scriptsize$\vdots$}};@)
%q = call ptr @__quantum__rt__qubit_allocate_array(i64 2) (@\qirline{qreg begin}@)(@\qirline{qreg end}\vspace{1pt}@)
%r = call ptr @__quantum__rt__array_create_1d(i32 8, i64 2) (@\qirline{creg begin}@)(@\qirline{creg end}\vspace{1pt}@)
%0 = call ptr @__quantum__rt__array_get_element_ptr_1d(ptr %q, i64 0) (@\qirline{h begin}@)
%q0 = load ptr, ptr %0, align 8
call void @__quantum__qis__h__body(ptr %q0) (@\qirline{h end}\vspace{1pt}@)
%1 = call ptr @__quantum__rt__array_get_element_ptr_1d(ptr %q, i64 1) (@\qirline{cx begin}@)
%q1 = load ptr, ptr %1, align 8
call void @__quantum__qis__cnot__body(ptr %q0, ptr %q1)  ; reuse %q0 (@\qirline{cx end}\vspace{1pt}@)
%r0 = call ptr @__quantum__qis__m__body(ptr %q0) (@\qirline{measure begin}@)
%2 = call ptr @__quantum__rt__array_get_element_ptr_1d(ptr %r, i64 0)
store ptr %2, ptr %r0, align 8
%r1 = call ptr @__quantum__qis__m__body(ptr %q1)
%3 = call ptr @__quantum__rt__array_get_element_ptr_1d(ptr %r, i64 1)
store ptr %3, ptr %r1, align 8 (@\qirline{measure end}@)
(@\hspace{1cm}\tikz[overlay] \node[anchor=base,yshift=-4pt] {{\scriptsize$\vdots$}};\vspace{1pt}@)
call void @__quantum__rt__qubit_release_array(ptr %q) ; release all resources
(@\hspace{1cm}\tikz[overlay] \node[anchor=base,yshift=-3pt] {{\scriptsize$\vdots$}};\vspace{-0pt}@)
            \end{lstlisting}
        \end{minipage}
        \hrule width \linewidth \relax
        \caption{
            The quantum \enquote{Hello World} program, \ie, a circuit to create a Bell state (bottom left), expressed in OpenQASM~2.0~(top left) and QIR with dynamically allocated qubits~(right).
            The corresponding lines of code in each fragment are linked.
            For more details, see also \cref{exp:qasm,exp:qir}.
        }
        \label{fig:code}
    \end{figure*}

    Overall, the described steps show that it does not take much to turn an existing simulator into a QIR runtime.
    The evaluation compares the new approach with the current method of parsing QASM files at runtime.
    Benchmark results indicate that both approaches yield similar performance, with no clear advantage in terms of runtime.
    However, when using QIR, one can take advantage of the entire compiler infrastructure that exists around \mbox{LLVM IR}.
    As a result, the execution of hybrid programs, where classical and quantum operations are interleaved arbitrarily, comes for free. 

    Throughout the article, we assume the reader to be familiar with the basics of quantum computing and quantum circuits.
    We refer the reader to the literature~\cite{nielsenQuantumComputationQuantum2010} for an in-depth introduction.

    \section[An Abridged History of Quantum Intermediate Representations]{An Abridged History of Quantum Intermediate Representations}\label{sec:history}

    First, this section briefly reviews the history of quantum intermediate representations and how quantum computations are typically represented.
    While certainly not exhaustive, these recollections provide a starting point for understanding the different approaches. %

    \subsection[OpenQASM 2: Quantum Computation approximately equals Quantum Circuit]{OpenQASM 2: \\Quantum Computation $\approx$ Quantum Circuit}\label{sec:openqasm2}

    The \emph{Open Quantum Assembly}~(OpenQASM) language was initially proposed in version~2 by IBM in 2017~\cite{crossOpenQuantumAssembly2017} to describe quantum experiments to be executed on their publicly available quantum computers.
    It is a low-level language that describes quantum computations in a straightforward manner, \ie, as an enumeration of quantum instructions.
    Additionally, it supports measurements, resets, (limited) feedback, and gate subroutines.
    Since its inception, OpenQASM has been widely adopted throughout the quantum computing community and is heavily used as a text-based exchange format for quantum circuits.

    \begin{example}
        \label{exp:qasm}
        The \enquote{Hello World} of quantum computing, the creation of a Bell state, is expressed in OpenQASM~2.0 on the top left in \cref{fig:code}.
        After loading the quantum (standard) library, the code declares a quantum register with two qubits and a classical register with two bits.
        Then, a Hadamard-gate is applied to the first qubit, followed by a CNOT-gate controlled by the first qubit and targeting the second qubit.
        Finally, both qubits are measured.
    \end{example}

    \subsection[OpenQASM 3: The Need for Hybrid Quantum Programs]{OpenQASM 3: \\The Need for Hybrid Quantum Programs}\label{sec:openqasm3}

    While OpenQASM~2 provided a great starting point, it has become apparent over time that some degree of classical logic and control flow, including conditionals and loops that might depend on the results of quantum measurements, is desirable for writing more complex quantum programs.
    For \mbox{near-term} applications, this allows to describe variational quantum algorithms, where the quantum circuit is part of a larger classical optimization loop.
    For \mbox{long-term} applications, classical feedback is essential for quantum error correction~\cite{campbellRoadsFaulttolerantUniversal2017,pehamAutomatedSynthesisFaultTolerant2024}.
    As a consequence, OpenQASM has been extended to version~3~\cite{crossOpenQASMBroaderDeeper2022}, which integrates classical logic and control flow.

    Fundamentally, OpenQASM started as a quantum assembly language and step-by-step evolved into a more general-purpose quantum programming language by adding classical elements on top.
    It builds on its strong adoption in the quantum computing community and the corresponding compiler infrastructure previously built around OpenQASM~2.
    At the same time, the extension to OpenQASM~3 requires implementing traditional compiler optimizations on top of the IR, such as loop unrolling, constant propagation, or constant folding to generate efficient executable code for the target hardware.
    Given how ubiquitous OpenQASM is in the quantum computing community, this extension is a natural and sensible step to enable more complex programs.
    However, it also requires the reimplementation of concepts that are already established and used for decades in classical compilers.
    This discrepancy has led to the development of QIR, which is discussed next.

    \subsection{QIR: Adding Quantum to the Classical}\label{sec:qir}

    In contrast to OpenQASM, the \emph{Quantum Intermediate Representation}~(QIR,~\cite{QIRSpec2021}) adopts a different strategy:
    It builds upon an established classical compilation infrastructure, augmenting it with quantum instructions.
    This approach allows QIR to utilize the existing classical compiler infrastructure, thereby inheriting its optimizations and transformations for classical code without additional effort.
    Specifically, QIR extends the classical IR used in the compiler framework \emph{LLVM}~\cite{lattnerLLVMCompilationFramework2004}.
    It uses this IR to implement various compiler optimizations on this IR before, eventually, compiling it to target-specific assembly for execution.
    As such, LLVM IR can express arbitrary classical program logic, including functions, conditionals, and loops.
    However, it cannot express quantum computations by itself.
    To this end, QIR, as initially proposed by \mbox{Microsoft}, defines a set of additional functions that can be used to express quantum computations with the LLVM IR.

    \begin{example}
        \label{exp:qir}
        The relevant lines of a QIR program that represent the same circuit as in \cref{exp:qasm} are shown on the right in \cref{fig:code}\footnote{Note, that we chose to use modern LLVM syntax with opaque pointers here, as opposed to the legacy syntax used in the initial QIR specification.}.
        Initially, an array \lstinline|%q| of two qubits and another array \lstinline|%r| to fit two measurement results are allocated.
        Note here, that the result type is an opaque pointer according to the QIR specification.
        Hence, we specify the element size to be 8 bytes to fit a pointer to the results.

        For the application of the Hadamard-gate, the first qubit is extracted from the qubit array.
        Since, \lstinline|%1| is the pointer to the first element in the array and not the qubit reference itself, we dereference this variable and store the qubit reference in \lstinline|%q|.
        Hereafter, the Hadamard-gate is applied to the qubit.
        In the next block of code, we also extract the second qubit \lstinline|%q1| and perform a CNOT-gate on both.
        Afterward, the state of both qubits is measured and stored in the initially allocated array \lstinline|%r|.
        Finally, after the results got processed by classical code or recorded in the output of the program, all resources are released---something that cannot be done explicitly in OpenQASM.
    \end{example}
    \begin{figure}[t]
        \centering
        \includegraphics[width=\linewidth]{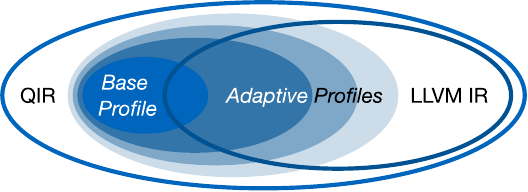}
        \caption{
            A Venn diagram showing the different profiles of QIR.
            The base profile is the most restrictive one and only uses very few instructions of the LLVM IR.
            QIR in its whole extent is a proper superset of LLVM IR.
        }
        \label{fig:qir profiles}%
    \end{figure}
    In order to speed up the adoption of QIR, multiple restrictions to QIR, so-called \emph{profiles} have been defined that limit the expressiveness of QIR\footnote{\url{https://github.com/qir-alliance/qir-spec/tree/main/specification/under_development/profiles} (visited on 03/03/2025)}.
    In its most restrictive form, the \emph{base profile} only allows a sequence of quantum instructions that ends with the measurement of all qubits, which effectively makes it very similar to OpenQASM~2.
    The more permissive \emph{adaptive profiles} allow the successive transition to fully support all features contained in LLVM IR.
    In its full generality, QIR is a proper superset of LLVM IR that builds on top of all the existing tooling and optimizations available for LLVM IR.
    For an illustration of the relation between the different profiles, see \cref{fig:qir profiles}.

    \section{Adopting QIR: Quo Vadis?}\label{sec:adoption}%

    The proposal of QIR is both logical and appealing: It makes sense to avoid reinventing the wheel and to build on decades of classical compilation expertise, especially, since quantum computations will inherently involve classical input and produce classical output.
    However, the adoption of QIR in the quantum computing software stack is a challenging task: What does it actually mean to adopt QIR in the stack?
    How do we transform these quantum-classical programs so that they can be executed?
    What about the actual execution?
    In the following, we discuss two fundamentally different directions of adopting QIR, where the first of both is subdivided into parsing and transforming QIR programs.

    \subsection{Parsing QIR Programs}\label{subsec:parsing}%

    In order for a quantum program to be executed, it must be transformed so that it complies with all the restrictions imposed by the hardware.
    Tools responsible for this transformation need to be able to accept the input quantum program and transform it accordingly~\cite{liTacklingQubitMapping2019,schoenbergerUsingBooleanSatisfiability2024,stadeAbstractModelEfficient2024,quetschlichMQTPredictorAutomatic2025,willeMQTHandbookSummary2024}.
    When using QIR, tools may either receive the program as a text file containing QIR code or as an in-memory representation of the program, such as an \emph{Abstract Syntax Tree}~(AST).
    The former approach requires a parser that can turn the text-based representation into some in-memory representation.
    This in-memory representation can either be the QIR AST, in which case LLVM can be used, or another custom and tool-specific IR.
    For a custom IR, the parser's (and, in fact, also the exporter's) complexity depends on the profile of QIR to be supported.

    \begin{example}
        \label{exp:parse}
        For the straight-line quantum programs, it suffices to iterate over the lines to construct an in-memory representation of the resulting quantum circuit.
        For example, when parsing the QIR program from \cref{exp:qir}, the parser would need to track the assignment of variables (\ie, \lstinline|%q|, \lstinline|%0|, \lstinline|%1|, \dots) to their values to infer the respective qubit that is passed to a quantum instruction.
        The instructions themselves can be matched with a simple pattern and corresponding actions can be taken to gradually build up the quantum circuit.
    \end{example}

    On the one hand, the advantage of a custom parser is that one can avoid the dependency on LLVM, which should be a carefully considered decision because of the substantial size and complexity of the module.
    LLVM as a dependency can significantly increase the build and maintenance overhead of a project.
    On the other hand, when not using LLVM, one has to reimplement all the optimizations and transformations that are already provided for LLVM~IR \enquote{for free}---similar to the ongoing development of tools for OpenQASM~3.
    Additionally, when using a custom IR, one is limited to the capabilities of that existing IR.
    If that IR is not expressive enough to capture all the details of a quantum program, the QIR program cannot be fully translated and, hence, the tool cannot be used on these programs.
    As a consequence, the IR would have to add support for representing and handling all the concepts that QIR introduces, which might be a further significant effort in addition to reimplementing the optimizations and transformations.

    \subsection{Transforming QIR Programs}\label{subsec:transforming}
    As a continuation of the first direction, after loading the quantum program, multiple passes in the software stack optimize and transform the program to meet all constraints imposed by the hardware.
    For the remainder of this section, we assume that the program is represented as a QIR AST in memory. %
    
    The support for QIR can, again, be realized in two different ways: Either the QIR is transformed directly, or the program is transpiled into another custom IR, transformed, and transpiled back into QIR.
    While the latter approach enables quick adoption of QIR, it carries the same deficits as parsing the text-based QIR file into a custom IR.
    For tools to support QIR directly, the AST of a QIR program must be handled and transformed.
    While this might be simple enough for the base profile, it becomes more complex for the adaptive profiles, which could have arbitrarily complex classical code in-between quantum instructions.
    This approach, however, has the advantage that the tool can directly take advantage of existing optimizations and transformations that are already implemented for QIR---the core motivation of an IR in a compiler.

    \begin{example}
        \label{exp:loops}
        One use case of classical \mbox{\textsc{for}-loops} is the application of a gate to multiple qubits.
        The following QIR snippet
        shows a simple \mbox{\textsc{for}-loop} that performs a Hadamard gate on the qubits with hardware addresses~\(0,\dots,9\).

        \begin{lstlisting}[language=qir]
    %i = alloca i32, align 4
    store i32 0, ptr %i, align 4              ; int i = 0
    br label for.header
for.header:
    %1 = load i32, ptr %i, align 4
    %cond = icmp slt i32 %1, 10               ; i < 10
    br i1 %cond, label %body, label %exit
body:
    %2 = load i32, ptr %i, align 4
    call void @__quantum__qis__h__body(
               ptr inttoptr (i64 %2 to ptr))  ; $\color{TUM_grey}\mathtt{h(q_i)}$
    %3 = load i32, ptr %i, align 4
    %4 = add nsw i32 %3, 1                    ; i++
    store i32 %4, ptr %i, align 4
    br label %for.header
exit: (@\dots@)
        \end{lstlisting}

        \noindent Since QIR builds on the LLVM infrastructure, it is straightforward to unroll any loops with statically known bounds in the QIR program.
        Hence, a quantum optimization pass does not have to handle the \mbox{\textsc{for}-loop}, but only sees the ten individual Hadamard gates that are applied to the qubits.
    \end{example}

    While the previous sections discussed handling and transforming QIR programs themselves, the question remains how these transformed programs are eventually executed.

    \subsection{Executing QIR Programs}
    In contrast to the type of support for QIR discussed so far, the execution of QIR programs is fundamentally different.
    Another direction of adopting QIR enables the execution of QIR programs and can either be seen as a complement to the previously discussed options or as an orthogonal approach.

    A file that contains LLVM IR bytecode can be executed directly with the \texttt{lli} tool provided by the LLVM project.
    This tool reads the bytecode and interprets it on the fly.
    Furthermore, the compiler \texttt{clang}, also part of the LLVM project, can compile the bytecode to machine code and emit an executable.

    Despite \texttt{clang} and \texttt{lli} being already able to parse QIR programs, when compiling or interpreting those files individually without any additional definitions, the tools will raise an error for every quantum instruction they encounter.
    The raised error message will point out that a definition for the respective function cannot be found.
    This is logical, as the quantum instructions are not part of the LLVM IR standard and, hence, \texttt{clang} and \texttt{lli} do not know how to handle them.
    To overcome this issue, and be able to compile QIR programs to executables, we have to provide the missing definitions for the quantum instructions.

    \begin{example}
        \label{exp:runtime}
        One example that takes this approach is the \emph{Catalyst Quantum Runtime} implemented by \mbox{Xanadu} that incorporates their classical quantum circuit simulator \emph{Lightning}~\cite{asadiHybridQuantumProgramming2024}, which is part of \emph{PennyLane}~\cite{bergholm2022pennylaneautomaticdifferentiationhybrid}.
        Every function, such as \lstinline|@__quantum__qis__h__body|, is implemented so that it modifies the internal state of the simulator to reflect the application of the respective gate.
        Note that these functions need not be implemented in LLVM IR, as this would be quite cumbersome.
        Instead, the definition of the QIR functions can be provided as an implementation (written in C/C++, Rust or other kinds of languages) of a C interface, which is then compiled to LLVM just as any other regular classical program code.
    \end{example}

    This approach is orthogonal to the previous two options as it only concerns the implementation of quantum instructions, while the actual program structure is handled by the runtime.
    To this end, this approach is perfectly suited for integrating classical simulation techniques with QIR, especially ones developed in a language that compiles to LLVM, such as C/C++ or Rust.
    The resulting binaries can be maximally optimized by the LLVM compiler infrastructure, which is a significant advantage.

    \section{Realization of a Classical QIR Runtime}\label{sec:simulator}

    In the following, we demonstrate how an existing quantum circuit simulator can be turned into a QIR runtime---illustrating the steps necessary to support QIR.
    For this demonstration, we use the quantum circuit simulator based on decision diagrams~\cite{willeToolsQuantumComputing2022,willeDecisionDiagramsQuantum2023} that is bundled with MQT Core~\cite{burgholzer2025MQTCore} as part of the \emph{Munich Quantum Toolkit}~(MQT~\cite{willeMQTHandbookSummary2024}).
    The resulting implementation of the QIR runtime is also publicly available as open-source software as part of \mbox{MQT Core} under \url{https://github.com/munich-quantum-toolkit/core}.

    \subsection{Implementing the Quantum Instructions}\label{subsec:simulator qis}
    The specification of QIR defines a couple of functions to deal with data structures used by the QIR runtime.
    For example, there are dedicated functions to create and manipulate arrays of qubits and measurement results (see also \cref{subsec:datastructures}).
    Additionally, the runtime provides special handling for strings and big integers.
    However, the core part of the QIR runtime is the quantum instruction set~(QIS).
    The name of those functions follows the following pattern: \lstinline|@__quantum__qis_<gate>__body|, where \lstinline|<gate>| is the name of the respective quantum gate.
    For example, the function that implements the rotation around the x-axis is declared as \lstinline|declare void @__quantum__qis__rx__body(ptr, double)|, where the first parameter is the qubit reference and the second one is the rotation angle.

    Any implementation of the specification must provide definitions for all declarations.
    The standard way to do that is to implement the C-API corresponding to the LLVM function signatures defined in the specification\footnote{The main reason for choosing C for the interface is that C has a well-defined and stable application binary interface (ABI), which allows the resulting object file to be properly linked with the LLVM IR code. This, in no way, precludes an implementation in higher-level languages such as C++ or Rust, as also demonstrated by this article.}.
    The following code fragment shows the definition of the QIR function corresponding to the RX gate.
    
    \begin{lstlisting}[language=qir++,label={lst:def rx}]
extern "C" {
void __quantum__qis__rx__body(Qubit* qubit, double phi) {
  auto& backend = QIR_DD_Backend::getInstance();
  backend.apply(qc::RX, phi, qubit);
}
} // extern "C"
    \end{lstlisting}
    The backend is implemented as a singleton class that is accessed via the static method \lstinline|QIR_DD_Backend::getInstance|.
    The first time this method is called, the singleton instance is created and initialized.
    Afterward, we can forward the application of any quantum gate to the backend and accordingly modify the quantum state represented in the runtime.
    Below, you see an excerpt of the implementation of the singleton class \lstinline|QIR_DD_Backend| that serves as a
    mediator between the QIR runtime and the simulator.

    \begin{lstlisting}[language=qir++,label={lst:sim-instance}]
class QIR_DD_Backend {
private:
  explicit QIR_DD_Backend(); // private constructor
public:
  /// Return the unique instance of the backend.
  static QIR_DD_Backend& getInstance() {
    static QIR_DD_Backend instance;
    return instance;
  }
// ... other attributes and methods
};
    \end{lstlisting}

    \subsection{Handling Dynamic Qubit Allocation}\label{subsec:simulator addresses}

    Qubits in QIR are represented as opaque pointers.
    Thus, a crucial part of the runtime is to translate between those opaque pointers and the qubit identifiers used in the simulator.
    The simulator under consideration uses consecutively numbered unsigned integers starting from $0$ to identify qubits in the underlying state.
    Hence, the opaque pointer values of qubits in the QIR program must be translated to the respective qubit identifiers in the simulator.
    QIR offers two qubit management strategies---either via dynamically allocating qubits or via static addressing.
    In the following, the management of dynamically allocated qubits is described.
    You can find more details on how to handle static addressing in~\cref{subsec:qubit addresses}.

    First, a definition for the QIR function to allocate a qubit needs to be provided.
    Note the surrounding \lstinline{extern "C"} declaration of the function.
    This is required because we must implement the C interface to finally be able to properly link the runtime with the QIR code.
    The \lstinline{extern "C"} declaration ensures that this part of the code is compiled as C code even though it is contained in a \texttt{.cpp} file.

    \begin{lstlisting}[language=qir++,label={lst:sim-alloc}]
extern "C" {
Qubit* __quantum__rt__qubit_allocate() {
  auto& backend = mqt::QIR_DD_Backend::getInstance();
  return backend.qAlloc();
}
} // extern "C"
    \end{lstlisting}

    The above function simply
    delegates the allocation of a new qubit to the runtime,
    which internally maintains a \lstinline|qRegister| hash map of unique addresses and qubit indices as follows.
    The dynamic qubit addresses are given out in increasing order starting from \lstinline|MIN_DYN_QUBIT_ADDRESS|.
    
    \begin{lstlisting}[language=qir++,label={lst:sim-alloc-impl}]
class QIR_DD_Backend {
private:
  std::unordered_map<const Qubit*, size_t> qRegister;
  uintptr_t currentMaxQubitAddress = MIN_DYN_QUBIT_ADDRESS;
  size_t currentMaxQubitId = 0;
...
public:
  auto qAlloc() -> Qubit* {
    auto* qubit =
        reinterpret_cast<Qubit*>(currentMaxQubitAddress++);
    qRegister.emplace(qubit, currentMaxQubitId++);
    return qubit;
  }
// ... other attributes and methods
};
    \end{lstlisting}

    The actual translation of an instruction's qubit addresses to the simulator's identifiers is delegated to the following function within the runtime.
    
    \begin{lstlisting}[language=qir++,label={lst:sim-addr-translation}]
template <size_t SIZE>
auto QIR_DD_Backend::translateAddresses(
    const std::array<Qubit*, SIZE>& qubits) 
    -> std::array<size_t, SIZE> {
  // transform opaque qubit pointers to qubit ids
  std::array<size_t, SIZE> qubitIds{};
  for(size_t i = 0; i < SIZE; ++i) {
    qubitIds[i] = qRegister.at(qubits[i]);
  }
  return qubitIds;
}
    \end{lstlisting}

    In particular, qubit addresses are never dereferenced in this setting, but only used as keys in the hash map.
    For an alternative approach, where qubit addresses point to valid memory locations and the qubit identifiers are stored in the memory, see~\cref{sec:conclusions}.

    No matter how the address translation is implemented, the runtime must cope with qubits that might be allocated and deallocated in the middle of the program execution.
    This dynamic allocation might not be necessary for QIR files that follow a certain profile that explicitly disallows allocation in the middle of the program.
    \Eg, the base profile prohibits dynamic qubit allocation entirely and only allows static addressing (see~\cref{subsec:qubit addresses}).
    However, the presented quantum runtime supports static and dynamic qubit addressing.
    As a consequence, the runtime must be able to dynamically adjust the size of the quantum state in the simulator to accommodate the qubits that are allocated during the program execution.

    For the chosen quantum circuit simulator that is based on decision diagrams, this enlargement of the quantum state is straightforward. %
    The following code fragment sketches this process for the implemented runtime: The input parameter is the maximum qubit identifier that is currently needed, \eg, in the application of a quantum gate. Based on that, the backend checks whether this identifier is already contained in the current state, and if not, accordingly extends the state with additional qubits.

    \begin{lstlisting}[language=qir++,label={lst:enlarge state}]
auto QIR_DD_Backend::enlargeState(uint64_t maxQubit) 
    -> void {
  if (maxQubit >= numQubitsInQState) {
    const auto d = maxQubit - numQubitsInQState + 1;
    numQubitsInQState += d;
    dd->resize(numQubitsInQState);
    (@\textcolor{TUM_grey}{[initalize new qubits in \(\ket{0}\)-state]}@)
  }
}
    \end{lstlisting}

    \subsection{Defining Additional Data-Structures}\label{subsec:datastructures}

    Besides the definition of qubits as pointers to an opaque type \lstinline|%Qubit|, the QIR specification also defines a couple of data structures that are used by the QIR runtime.
    This subsection illustrates their realization in the implemented runtime using the example of the \lstinline|%Array| type.
    To not leave the array type incomplete, the backend provides the following definition with attributes that are necessary to implement all QIR runtime functions related to arrays.

    \begin{lstlisting}[language=qir++,label={lst:array def}]
using Array = struct ArrayImpl {
  int32_t refcount;          // reference count
  int32_t aliasCount;        // count of aliases
  std::vector<int8_t> data;  // vector of bytes
  int64_t elementSize;       // number of bytes per elem.
};
    \end{lstlisting}

    The QIR's specification enforces the principle of reference counting on classical resources like arrays. %
    For arrays specifically, alias counting is employed to keep track of read-only references to the array---the handling of those is omitted here for brevity.

    The life-cycle of a classical resource in QIR, and in particular of an array includes its creation, access, and release.
    First, the code snippet below provides the implementation of a one-dimensional array creation for a given element size \lstinline|s| in number of bytes and the total number of elements \lstinline|n|.

    \begin{lstlisting}[language=qir++,label={lst:array create}]
Array* __quantum__rt__array_create_1d(
    int32_t s, int64_t n) {  // s: element's size
  return new Array{1, 0, std::vector<int8_t>(s * n), s};
}
    \end{lstlisting}

    During program execution data should be written and read from the array.
    To this end, the function below returns a pointer to the \(i\)-th element of the array.
    Using this pointer, the data in the corresponding element can be accessed or modified.

    \begin{lstlisting}[language=qir++,label={lst:array get}]
int8_t* __quantum__rt__array_get_element_ptr_1d(
    Array* arr, int64_t i) {
  (@\textcolor{TUM_grey}{[perform bounds checking]}@)
  return &arr->data[arr->elementSize * i];
}
    \end{lstlisting}

    At the end, when the array is no longer needed, the associated memory should be freed to make space for other resources.
    The deallocation of classical resources is part of the reference counting mechanism in QIR.
    Hence, the function to update the reference count, frees the associated memory when the reference count decreases to zero.

    \begin{lstlisting}[language=qir++,label={lst:array release}]
void __quantum__rt__array_update_reference_count(
    Array* arr, int32_t k) {
  if (arr == nullptr) return;
  arr->refcount += k;  // k can also be negative (!)
  if (arr->refcount == 0) delete arr;
}
    \end{lstlisting}

    \section{Demonstrating the Runtime's Functionality}\label{sec:evaluation}

    The explanations and code snippets above have, hopefully, given a rough overview of what it takes to realize a QIR runtime based on an existing quantum circuit simulator. %
    This section complements the description of the implementation by providing the results of running multiple benchmarks with the implemented runtime.
    This achievement demonstrates our ability to streamline the process of connecting quantum and classical systems without reinventing the wheel.

    For the evaluation, we considered five different families of quantum circuits from the MQT Bench quantum circuit benchmark suite~\cite{quetschlichMQTBenchBenchmarkingSoftware2023}.
    Those correspond to the first five families listed in \cref{tab:results} and can be realized using the QIR base profile.
    Additionally, we created three other sets of benchmark circuits from the MQT Core library~\cite{willeMQTHandbookSummary2024}.
    Those last three circuit families in \cref{tab:results} all make use of QIR features defined as part of the adaptive profile.
    More precisely, they employ intermediate measurements, reuse already measured qubits, and, in case of the \texttt{qft-iter} and \texttt{qpeexact-iter} circuits, furthermore contain quantum operations that are conditioned on previous measurement results.

    At the time of writing, neither MQT Bench nor MQT Core directly produce QIR code.
    Thus, all circuits were initially created as OpenQASM 3 programs and subsequently translated to equivalent QIR programs.
    During this process, we created two versions of the QIR files: one that uses dynamically allocated qubits as described in the implementation section, \ie, \cref{subsec:simulator addresses}, and one that uses statically addressed qubits (see \cref{subsec:qubit addresses}).

    Before the QIR files can be executed, they must be compiled to an executable.
    For this, every QIR file is linked to the implemented QIR runtime separately and compiled to an executable.
    Those executables can then be invoked, and their execution time can be measured.
    For a comparison between the execution of the QIR files and the existing MQT simulator, we employed the default simulation capabilities of the MQT to parse and simulate OpenQASM 3 files.

    During the evaluation we collected the following metrics:
    \begin{enumerate}
        \item\label{itm:comp} \textbf{Compilation Times}
        \begin{enumerate}
            \item\label{itm:app} of the application to parse and simulate OpenQASM 3 files
            \item\label{itm:lib} of the library containing the QIR runtime (without a specific QIR file)
            \item of every QIR file using the already compiled library from \cref{itm:lib}
        \end{enumerate}
        \item\label{itm:file} \textbf{File Sizes}
        \begin{enumerate}
            \item of the application from \cref{itm:app}
            \item of the OpenQASM files parsed by the application
            \item of the executables compiled from the QIR files
        \end{enumerate}
        \item\label{itm:exec} \textbf{Execution Times}
        \begin{enumerate}
            \item of the application to parse and simulate OpenQASM 3 files
            \item of the QIR executables using the QIR runtime
        \end{enumerate}
    \end{enumerate}

    Note that the application and the library only need to be compiled once, which is why we record the times of \cref{itm:app,itm:lib} only once.
    The results of all experiments, which were executed on a machine with an Apple M3 chip, 16\,GB of RAM, and \texttt{clang} version 16.0.0, are summarized in \cref{tab:results}.

    \begin{table*}
        \caption{Evaluation Results}
        \label{tab:results}
        \small
        \rule{\linewidth}{\heavyrulewidth}
        QASM simulator executable file size \faIcon{file}: \SI{744}{\kilo\byte} and compilation time \faIcon{hourglass}: \SI{10.496}{\second} \hfill QIR runtime library compilation time \faIcon{hourglass}: \SI{3.028}{\second}\\[4pt]
        \def\dynQASM{\multicolumn{4}{c}{\textbf{Dynamically Parsing QASM}}}
        \def\compiledQIR{\multicolumn{7}{c}{\textbf{Executing Compiled QIR}}}
        \def\staticQIR{\multicolumn{3}{c}{\textbf{Static Addressing}}}
        \def\dynQIR{\multicolumn{3}{c}{\textbf{Dynamic Addressing}}}
        \def\benchmark{\textbf{Benchmark}}
        \def\nQubits{{\#Qubits}}
        \def\compTime{{\faIcon{hourglass} [\si{\second}]}}
        \def\execTime{{\faIcon{clock} [\si{\second}]}}
        \def\fileSize{{\faIcon{file} [\si{\kilo\byte}]}}
        \begin{tabularx}{\linewidth}{%
            r%
            S[table-format=4.0]%
            X@{\hspace*{2ex}}%
            S[table-format=6.0]%
            S[table-format=2.3]%
            X%
            S[table-format=1.3]%
            S[table-format=6.0]%
            S[table-format=2.3]%
            X%
            S[table-format=1.3]%
            S[table-format=6.0]%
            S[table-format=2.3]%
        }
            \toprule
            & & \dynQASM & \compiledQIR \\
            & & & & & & \staticQIR & & \dynQIR \\
            \benchmark             & \nQubits & & \fileSize & \execTime & & \compTime & \fileSize & \execTime & & \compTime & \fileSize & \execTime  \\
            \midrule%
            \texttt{ghz}           & 256      & & 13        & 0.008     & & 3.828     & 324       & 0.013     & & 3.846     & 346       & 0.019     \\
            & 512      & & 26        & 0.028     & & 3.864     & 349       & 0.045     & & 3.875     & 371       & 0.089     \\
            & 1024     & & 53        & 0.141     & & 3.911     & 398       & 0.201     & & 3.863     & 420       & 0.273     \\
            & 2048     & & 111       & 0.736     & & 3.973     & 480       & 0.980     & & 3.937     & 535       & 2.340     \\
            & 4096     & & 228       & 4.412     & & 4.399     & 662       & 5.675     & & 4.013     & 931       & 6.807     \\[1ex]
            \texttt{qft}           & 30       & & 12        & 0.003     & & 3.822     & 317       & 0.002     & & 3.849     & 339       & 0.002     \\
            & 35       & & 17        & 0.004     & & 3.873     & 317       & 0.004     & & 3.843     & 339       & 0.004     \\
            & 40       & & 22        & 0.005     & & 3.838     & 317       & 0.004     & & 3.848     & 340       & 0.005     \\
            & 45       & & 27        & 0.006     & & 3.899     & 318       & 0.005     & & 3.866     & 340       & 0.005     \\[1ex]
            \texttt{qpeexact}      & 30       & & 13        & 0.003     & & 3.825     & 317       & 0.003     & & 3.858     & 340       & 0.002     \\
            & 35       & & 18        & 0.007     & & 3.934     & 318       & 0.003     & & 3.838     & 340       & 0.004     \\
            & 40       & & 23        & 0.006     & & 3.823     & 318       & 0.004     & & 3.867     & 340       & 0.004     \\
            & 45       & & 29        & 0.208     & & 3.919     & 334       & 0.005     & & 4.108     & 340       & 0.006     \\[1ex]
            \texttt{random}        & 14       & & 11        & 0.193     & & 4.175     & 339       & 0.218     & & 3.941     & 344       & 0.216     \\
            & 15       & & 13        & 0.713     & & 3.948     & 339       & 0.736     & & 3.983     & 344       & 0.761     \\
            & 16       & & 14        & 2.088     & & 3.968     & 339       & 2.110     & & 3.917     & 361       & 2.151     \\
            & 17       & & 15        & 18.315    & & 3.905     & 340       & 17.703    & & 4.302     & 362       & 17.430    \\
            & 18       & & 18        & 95.679    & & 3.913     & 340       & 89.789    & & 4.011     & 362       & 90.568    \\[1ex]
            \texttt{wstate}        & 256      & & 33        & 0.042     & & 3.863     & 341       & 0.049     & & 3.850     & 363       & 0.057     \\
            & 512      & & 67        & 0.165     & & 3.892     & 382       & 0.237     & & 4.075     & 404       & 0.259     \\
            & 1024     & & 136       & 0.802     & & 4.274     & 448       & 1.004     & & 3.938     & 486       & 1.023     \\
            & 2048     & & 282       & 4.544     & & 4.283     & 613       & 5.397     & & 4.033     & 668       & 5.709     \\
            & 4096     & & 574       & 30.504    & & 5.599     & 920       & 34.191    & & 4.316     & 1295      & 36.063    \\[1ex]
            \texttt{bv-iter}       & 256      & & 15        & 0.002     & & 3.830     & 346       & 0.002     & & 3.962     & 347       & 0.002     \\
            & 512      & & 29        & 0.003     & & 3.879     & 371       & 0.002     & & 3.976     & 371       & 0.002     \\
            & 1024     & & 58        & 0.006     & & 3.952     & 420       & 0.003     & & 3.910     & 437       & 0.002     \\
            & 2048     & & 117       & 0.011     & & 4.377     & 535       & 0.004     & & 3.863     & 552       & 0.004     \\
            & 4096     & & 236       & 0.020     & & 6.923     & 750       & 0.007     & & 4.117     & 866       & 0.006     \\[1ex]
            \texttt{qft-iter}      & 30       & & 22        & 0.002     & & 3.839     & 338       & 0.002     & & 3.881     & 339       & 0.001     \\
            & 35       & & 30        & 0.002     & & 4.265     & 338       & 0.002     & & 3.827     & 339       & 0.001     \\
            & 40       & & 40        & 0.003     & & 4.192     & 338       & 0.002     & & 3.871     & 355       & 0.002     \\
            & 45       & & 51        & 0.003     & & 3.828     & 355       & 0.001     & & 3.865     & 356       & 0.002     \\[1ex]
            \texttt{qpeexact-iter} & 30       & & 24        & 0.002     & & 3.822     & 340       & 0.002     & & 3.859     & 341       & 0.002     \\
            & 35       & & 32        & 0.003     & & 3.822     & 357       & 0.002     & & 3.839     & 358       & 0.001     \\
            & 40       & & 42        & 0.003     & & 3.840     & 357       & 0.002     & & 3.844     & 358       & 0.001     \\
            & 45       & & 53        & 0.004     & & 3.883     & 357       & 0.002     & & 3.860     & 358       & 0.002     \\
            \bottomrule
        \end{tabularx}\\[4pt]
        \small
        \faIcon{hourglass}: Compilation times in seconds (see \cref{itm:comp})\hfill
        \faIcon{file}: Files Sizes in kilobytes (see \cref{itm:file})\hfill
        \faIcon{clock}: Execution Times seconds (see \cref{itm:exec})
    \end{table*}

    At the top of the table, we provide the file sizes and compilation times of the QASM simulator application together with the compilation time for the QIR runtime library.
    Note that the compilation of the simulator executable takes more time than the compilation of the QIR runtime library together with each QIR file.
    However, the application only needs to be compiled once and can then be used to parse and simulate multiple OpenQASM files without recompilation, while a separate executable has to be compiled per QIR file.

    Regarding the results of the individual circuits, we first take a look at the resulting file sizes.
    For the simulation of QASM files, the application together with the QASM file is required to run the respective circuit.
    Hence, we compare the sum of the file sizes of the application and the QASM file with the size of the compiled QIR executables.
    It stands out that for smaller circuits, the compiled QIR files are significantly smaller.
    However, for larger circuits, the file size of the QIR executables increases faster than the file size of the QASM files.
    Nevertheless, for all circuits in the evaluation, the size of the QIR executables remains below the size of the application and the QASM file together.

    With a look at the execution times, all three different settings lead to similar results\footnote{
        One may notice that the qubit ranges and the execution time scales between the individual circuit families differ wildly.
        This is a testament to the properties of decision diagrams, which are used as the data structure for the simulation.
        Similar to tensor networks, decision diagrams allow representing certain problem classes much more efficiently and compactly than an array-based statevector representation by exploiting structure in the underlying representation.
        At the same time, they incur a significant overhead for problem classes that do not offer any benefit for the decision diagram structure as, e.g., seen in the results for the \texttt{random} circuit family~\cite{grurlArraysVsDecision2020}.
    }.
    This is hardly surprising, as all of the approaches are carrying out identical quantum computations, and the majority of the execution time is spent applying the respective operations to the current state.
    Hence, the overhead from parsing the file within the application to simulate QASM files is negligible.
    The execution times of the QIR files using dynamic and static qubit addresses also do not differ significantly.
    
    In summary, those results manifest that with manageable effort, an existing quantum circuit simulator can be turned into a QIR runtime while maintaining the same performance.
    This opportunity opens up a way for fast adoption of the QIR standard with tools that were previously developed to work with QASM files or similar intermediate representations.
    On top of that, one unlocks the ability to execute arbitrary hybrid classical-quantum programs such as quantum error correction protocols with classical feedback loops or dynamic quantum circuits as demonstrated with the iterative versions of the Bernstein-Vazirani algorithm, the Quantum Fourier Transform, and the Quantum Phase Estimation algorithm in \cref{tab:results}.

    \section{Challenges in Compiling QIR Programs}\label{sec:challenges}

    The integration of a quantum IR like QIR into a compiler infrastructure is challenging and comes with a number of open research questions.
    Here, we detail two of those challenges. %

    \subsection{Static and Dynamic Qubit Addresses}\label{subsec:qubit addresses}

    QIR provides the opaque pointer type \lstinline|%Qubit*| for addressing individual qubits.
    In \cref{exp:qir}, the qubits were addressed dynamically by first allocating the respective qubits and retrieving dynamic qubit addresses that later can be used as arguments to quantum instructions.
    In contrast to dynamic addresses, there is also the possibility to address qubits statically.

    \begin{example}
        \label{exp:qir static}
        When addressing the qubits statically instead of dynamically, the circuit from \cref{fig:code} can be rewritten in QIR in the following snippet.
        Especially, the lines for allocating the qubits disappear.

        \begin{lstlisting}[language=qir]
(@\hspace{1cm}\tikz[overlay] \node[anchor=base,yshift=-1.5pt] {{\scriptsize$\vdots$}};@)
call void @__quantum__qis__h__body(ptr null)
call void @__quantum__qis__cnot__body(ptr null,
    ptr inttoptr (i64 1 to ptr))
call void @__quantum__qis__mz__body(ptr null, ptr null)
call void @__quantum__qis__mz__body(ptr inttoptr
    (i64 1 to ptr), ptr inttoptr (i64 1 to ptr))
(@\hspace{1cm}\tikz[overlay] \node[anchor=base,yshift=-3pt] {{\scriptsize$\vdots$}};\vspace{-1pt}@)
        \end{lstlisting}
    \end{example}

    This type of addressing qubits can be useful when compiling a quantum program to an executable very close to hardware.
    In the end, the hardware only has a fixed number of qubits, and the compiler must ensure that the program does not exceed this number.
    So while a quantum program can be written with dynamic qubit addresses, the compiler must at some point assign the program's qubits to the hardware's qubits---a process very similar to register allocation in classical compilers.
    At the moment, this distinction between static and dynamic qubit addresses is not yet fully established in the QIR community, but it is an important aspect to consider when integrating QIR into a compiler infrastructure.

    In the context of implementing a QIR runtime for a quantum circuit simulator---as outlined above---dynamic qubit addresses are the preferred way to address qubits.
    Most simulators support a variable number of qubits; the maximum feasible number might even depend on the input circuit itself, \eg, on the degree of entanglement that is created during its execution.
    By allocating the qubits on demand, one can efficiently scale the simulated state to the required size and save memory if no large state is needed.

    To support static qubit addresses, the runtime would have to infer the number of qubits required for the simulation somehow.
    According to the examples in the QIR specification, every function in a QIR file has an attribute denoting the number of required qubits by this function.
    Those attributes can be accessed by compilation passes in the process of compiling the source code into an executable.
    However, in our setting, where we just provide the definitions of the QIR runtime functions as a C/C++ implementation, we cannot access those attributes from the QIR files.
    One can still allocate qubits on the fly when encountering a new qubit address that is not yet part of the simulated quantum state (similar to how this is handled in the runtime implementation evaluated in~\cref{sec:evaluation}).

    \subsection{Hybrid Classical-Quantum Computing}\label{subsec:hybrid}
    As QIR is a superset of LLVM IR, arbitrarily complex classical computations can be expressed in LLVM IR.
    Those heavy classical computations are preferably executed on dedicated classical hardware rather than on the classical co-processor of a quantum computer---if that is even feasible:
    The classical co-processor of quantum computers must be very fast, and special-purpose hardware like FPGAs or ASICs are employed in many such cases.
    These are incapable of executing arbitrary classical code, which is also not their purpose.
    Consequently, the question of how to decide which part of the code should be executed on the classical hardware and which part on the quantum hardware naturally arises for a hybrid classical-quantum programs that contain quantum as well as classical instructions.

    Especially in the realm of error correction, where conditional gate applications based on intermediate measurements must be performed on the quantum computer to ensure low latency, the distinction is more complicated than just offloading only the quantum instructions.
    At the same time, it must be ensured that the classical code offloaded to the quantum hardware can be executed in the required time frame to uphold the coherence of the qubits.
    Hence, as long as quantum computers cannot achieve arbitrary coherence of the qubits, there will always be programs that describe an infeasible execution and must be rejected.

    \section{Conclusions}\label{sec:conclusions}
    QIR is one way to bridge the gap between purely classical and purely quantum computations.
    As an extension of LLVM IR, it builds on decades of compiler infrastructure and optimizations.
    As outlined in this article, the adoption of QIR can be achieved with less effort than it might seem at first glance.
    However, some open research questions remain to be addressed in the future, especially in the context of hybrid classical-quantum programs.
    One framework for exploring solutions to these questions is the \emph{Multi-Level Intermediate Representation}~(MLIR,~\cite{mlir}), which is a natural choice for the next step in the evolution of QIR.

    \subsection*{Acknowledgements}\label{sec:ack}
    The authors acknowledge funding from the European Research Council (ERC) under the European Union’s Horizon 2020 research and innovation program (grant agreement No. 101001318), and the Munich Quantum Valley (MQV), which is supported by the Bavarian state government with funds from the Hightech Agenda Bayern Plus.

    \renewcommand*{\bibfont}{\small} %
    \printbibliography

\end{document}

\typeout{get arXiv to do 4 passes: Label(s) may have changed. Rerun}